# Designable spectrometer-free index sensing using plasmonic Doppler gratings

Fan-Cheng Lin,[†,#] Kel-Meng See,[†,#] Lei Ouyang,[‡,§] You-Xin Huang,[†] Yi-Ju Chen,[‡] Jürgen Popp[‡,§] and Jer-Shing Huang[*,†,‡,⊥,‖]

[†]Department of Chemistry, National Tsing Hua University, Hsinchu 30013, Taiwan

[‡]Leibniz Institute of Photonic Technology, Albert-Einstein Straße 9, D-07745 Jena, Germany

[§]Institute of Physical Chemistry and Abbe Center of Photonics, Friedrich-Schiller-Universität Jena, Helmholtzweg 4, D-07743 Jena, Germany

[⊥]Research Center for Applied Sciences, Academia Sinica, 128 Sec. 2, Academia Road, Nankang District, Taipei 11529, Taiwan

[‖]Department of Electrophysics, National Chiao Tung University, Hsinchu 30010, Taiwan

**ABSTRACT:** Typical nanoparticle-based plasmonic index sensors detect the spectral shift of localized surface plasmon resonance (LSPR) upon the change of environmental index. Therefore, they require broadband illumination and spectrometers. The sensitivity and flexibility of nanoparticle-based index sensors are usually limited because LSPR peaks are usually broad and the spectral position cannot be freely designed. Here, we present a fully designable index sensing platform using plasmonic Doppler gratings (PDGs), which provide broadband and azimuthal angle dependent grating periodicity. Different from LSPR sensors, PDG index sensors are based on the momentum matching between photons and surface plasmons via the lattice momentum of the grating. Therefore, index change is translated into the variation of in-plane azimuthal angle for photon-to-plasmon coupling, which manifests as directly observable dark bands in the reflection image. The PDG can be freely designed to optimally match the range of index variation for specific applications. In this work, we demonstrate PDG index sensors for large (n = 1.00~1.52) and small index variation (n = 1.3330~1.3650). The tiny and nonlinear index change of water-ethanol mixture has been clearly observed and accurately quantified. Since the PDG is a dispersive device, it enables on-site and single-color index sensing without a spectrometer and provides a promising spectroscopic platform for on-chip analytical applications.

**Introduction** Plasmonic nanoparticle-based sensors have attracted tremendous attention.[1-4] The working principle is based on detecting the spectral shift of the localized surface plasmon resonance (LSPR). Since LSPR is determined by the particle shape, materials and the surrounding medium[5,6], the change in the environmental index can be quantified by measuring the spectral shift of LSPR. This kind of nanoparticle-based plasmonic sensors has found many applications, including heavy metal pollutant detection[7,8], hydrogen sensing[9,10], disease diagnosis[11,12], chemical reaction monitoring[13,14] and index sensing[15,16]. However, the sensitivity of nanoparticle index sensors based on LSPR is typically limited by the broad bandwidth of the LSPR peaks, which is mainly due to the intrinsically low quality factor of plasmonic resonator.[17] To have reasonable sensitivity, LSPR based index sensors either require very large spectral shift upon particle aggregation[7,18,19] or rely on sharp spectral features, e.g. the steep edge of Fano-like resonance of specially engineered nanostructures.[20-22] Moreover, the LSPR of single nanoparticles cannot be freely designed because LSPR is an intrinsic property of the nanoparticle. Therefore, the sensitivity, i.e. the spectral shift per refractive index unit (RIU), can hardly be optimally tuned for different applications. From experimental point of view, the major drawback of LSPR based index sensors is that they require broadband illumination and a spectrometer to perform spectroscopic analysis. Although there are colorimetric sensors developed for bare-eye detection, they are more suitable for qualitative analysis [23-25] rather than quantitative one, for which high-performance spectrometers are still necessary to detect the small shift of the broad LSPR peaks.

To address these issues, we propose using a plasmonic Doppler grating (PDG) as a designable index sensor, which is capable of on-site, single-color and spectrometer-free sensing. The working principle of PDGs is based on the coupling between free-space photon and surface plasmons (SPs) via the azimuthal angle dependent gratings, instead of LSPR. Typically, direct excitation of SPs by free space photons is not possible because SPs possess larger in-plane momentum than the corresponding pho-tons. Several schemes have been developed to excite SPs at metal/dielectric boundaries, including Kretschmann- and Otto-configurations[26,27], nonlinear wave-mixing[28, 29] and metallic gratings[30-32]. Among these strategies, metallic gratings have been extensively used and investigated because of their simplicity and well understood coupling mechanism.[32-37] A grating can effectively excite SPs because it provides lattice momentum to fulfill the momentum conservation condition (Fig. 1a)[30]

$$\frac{2\pi}{\lambda_0} n_d \sin\alpha + \frac{2m\pi}{P} = \pm \frac{2\pi}{\lambda_0}\sqrt{\frac{\varepsilon_m \cdot n_d^2}{\varepsilon_m + n_d^2}}. \qquad (1)$$

Here, α is the incident angle, $\lambda_o$ is the vacuum wavelength of the incident light, $\varepsilon_m$ is the permittivity of material, $P$ is the periodicity of the grating, $n_d$ is the refractive index of the surrounding dielectric medium and $m$ is the resonant order. Eq. (1) is the base of index sensing using plasmonic gratings as it links the refractive index of the surrounding medium ($n_d$) to the grating periodicity ($P$). Metallic gratings have been used for index sensing.[38-40] Recently, chirped gratings have been demonstrated to offer continuously varying periods in one direction.[41-43] As described in our previous work,[30] our Doppler grating design also provides continuously varying periodicity depending on the azimuthal angle. The azimuthal angle dependent grating periodicity can be calculated by Eq. (2)

$$P(\varphi) = \pm d\cos\varphi + \sqrt{(d^2 \cos 2\varphi + 2\Delta r^2 - d^2)/2}. \qquad (2)$$

Here, $\varphi$ is the in-plane azimuthal angle, at which free space photons effectively excite SPs. $\Delta r$ is the radius increment and $d$ is the displacement of the ring center. They are the two most important design parameters for a PDG because they determine the span and the central wavelength of the spectral window. By choosing these two design parameters, a PDG can be easily designed for optimal sensing performance for large index change or small index change. Inserting Eq. (2) into (1), the coupling wavelength can be expressed as

$$\lambda_0 = \frac{\pm d\cos\varphi + \sqrt{(d^2 \cos 2\varphi + 2\Delta r^2 - d^2)/2}}{m}\left(\sqrt{\frac{\varepsilon_m \cdot n_d^2}{\varepsilon_m + n_d^2}} - n_d \sin\alpha\right). \qquad (3)$$

Equation (3) links the refractive index of the surrounding medium ($n_d$) to the in-plane azimuthal angle ($\varphi$) for PDG index sensing. For a well-designed PDG structure (fixed $\Delta r$, $d$ and $\varepsilon_m$) illuminated by single-color light source impinging at a specific incident angle (fixed $\lambda_0$ and $\alpha$), photon-to-plasmon coupling only happens effectively along specific azimuthal angles determined by the refractive index of surrounding medium. One unique feature of PDGs is that they offer varying periodicity chirped in the tangential direction perpendicular to the grating's wave vector, whereas linear chirped gratings provide varying periodicity chirped in the direction parallel to the grating's wave vector. Therefore, spectral dispersion due to conventional grating effect and the periodicity chirp can be unambiguously separated by PDGs but not by linear chirped gratings. Also, PDGs provide larger number of grooves for each periodicity than linear chirped gratings and thus better grating efficiency. In an experiment, the change of the in-coupling azimuthal angle can be easily observed because photon-to-plasmon coupling would manifest as dark bands in the reflection images. In the following, we demonstrate index sensing with reflection type PDGs (Fig. 1b) optimally designed for large range ($n_d$ = 1.00 to 1.52) and small range ($n_d$ = 1.3330 to 1.3650) index sensing. Experimental results are verified with the analytical model described by Eq. (3) .

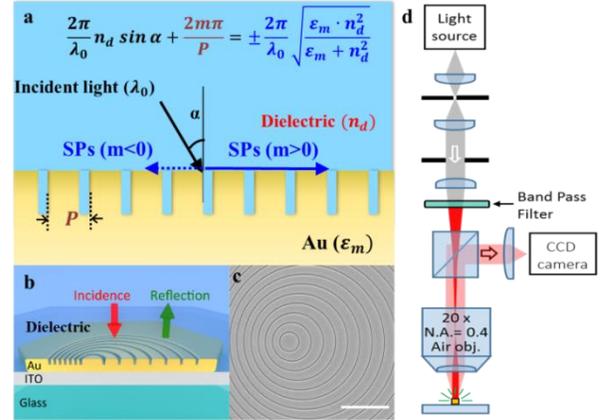

Figure 1. (a) A schematic illustrating photon-to-plasmon coupling via plasmonic grating. The momentum matching condition described in Eq. (1) governs the excitation of SPs, which can propagate into opposite directions depending on the resonance order $m$. (b) A schematic showing the structure of reflection type PDG index sensor and the configuration of optical characterization. (c) SEM image of a PDG fabricated on the surface of a single-crystalline gold flake ($\Delta r$ = 500 nm, $d$ = 140 nm) for large range index sensing. The scale bar is 2.5 μm. (d) Optical setup for near-normal incident reflection imaging.

**Method and materials**
**Nanofabrication** PDGs were fabricated by applying gallium focused-ion beam (FIB) milling (Helios Nanolab 600i System, FEI Company) to create grooves on the surface of chemically grown ultrasmooth single-crystalline gold flakes[30, 44] on a cover glass coated with an indium-tin oxide (ITO) transparent conductive layer (Fig. 1b). The ITO layer (thickness = 40 nm) helps avoid charging effect during FIB milling and scanning electron microscope (SEM) imaging. The beam current and the acceleration voltage of the FIB are 2.1 pA and 30 kV, respectively. The flakes are chemically stable and grain boundary free, making them an ideal substrate for the fabrication of high-definition plasmonic nanostructures.[30, 44, 45] The ultrasmooth surface of the flakes is important because it greatly suppresses the background noise from the scattering of SPs by surface roughness. This enhances the signal-to-noise ratio of the reflection images and facilitates the quantitative analysis of intensity angle distribution. Since the targeted media only approach the PDG from one side, only shallow PDG grooves (depth < 100 nm) were fabricated by FIB on the surface of thick gold flakes (thickness > 5 μm). Figure 1c shows the SEM image of the PDG for wide refractive index sensing ($n_d$ = 1.00 to 1.52). If thin gold flakes were used (thickness< 200 nm), FIB could mill through the flakes and create PDGs on both metal/dielectric interfaces, i.e. medium/gold and gold/substrate. This would allow photon-to-plasmon coupling on both interfaces and enable more functionali-ties, with the cost of more difficulty in quantitative analysis.[30, 34]

**Optical measurement** The reflection images of the fabricated PDGs were recorded by a home-made optical microscope. Figure 1b illustrates the optical characterization in reflection mode. The incident illumination is a near-normally incident white light. Since plasmonic gratings best respond to incident light with polarization perpendicular to the grating grooves, radially polarized light would be ideal for illumination. This requires, however, the center of the radially polarized illumination to be well aligned to the center of the zeroth ring of the PDG. Misalignment can lead to distortion of the scattering intensity distribution and lead to systematic bias. In this work, we have used unpolarized plane-wave illumination to reduce the complexity of optics and alignment. This is advantageous to practical applications of PDGs. The price to pay is the increase of background noise due to the reflection of light with unwanted polarization. Nevertheless, the intensity contrast due to photon-to-plasmon coupling is large enough to clearly reveal the position of dark bands for further quantitative analysis on the intensity angle profiles (Fig. 2a). Possible ellipticity of the unpolarized illumination introduced by optics in the illumination beam path has been taken into consideration by the polarization parameter η in the fitting formula (Supporting Information). The reflection image is collected by an air objective with low magnification (20×) and small numerical aperture (NA = 0.4), as shown in Figure 1d. Various transparent immersion media with different refractive indexes were introduced onto the surface of PDG and covered by a second cover glass. The used media include air ($n_d$ = 1.00), water ($n_d$ = 1.33), ethanol ($n_d$ = 1.36), tert-butanol ($n_d$ = 1.39), ethylene glycol ($n_d$ = 1.43) and microscope index matching oil ($n_d$ = 1.52). For large range index sensing ($n_d$ = 1.00 ~ 1.52), bandpass filters with 40 nm bandwidth (FKB-VIS-40, Thorlabs) were used to obtain narrow-band illumination centered at 550 nm and 650 nm. For small range index sensing on water-ethanol mixtures ($n_d$ = 1.3330~1.3650), a 632.8 nm laser line filter was used (bandwidth = 3 nm, FL632.8-3, Thorlabs). To avoid contamination from residual immersion media, after each immersion and sensing process, the PDGs were rinsed with ethanol and deionized water and blown dry with clean pressurized nitrogen.

**Results and discussions**
The angle distribution of grating periodicity and thus the sensitivity can be freely designed by choosing suitable $\Delta r$ and $d$. From azimuthal angle 0° to 180°, the periodicity changes continuously from the maximum $\Delta r + d$ to the minimum $\Delta r - d$. For example, for large index change ($n_d$ = 1.00 to 1.52), the PDG has been designed to possess grating periodicity from 360 nm to 640 nm by choosing $\Delta r$ = 500 nm and $d$ = 140 nm. For sensing the small index variation of the water-ethanol mixture ($n_d$ = 1.3330 to 1.3650), the PDG has been designed to "zoom-in" to this range by choosing $\Delta r$ = 390 nm and $d$ = 40 nm. The possibility to freely design and optimize the sensitivity for specific applications makes PDG a convenient and unique index sensing platform.

Figure 2a shows the reflection images of the PDG index sensor optimized for large index change. Under the coverage of six different dielectric media, the surrounding refractive index changes from 1.00 to 1.52. As can be seen in the full-color images (left column in Fig. 2a), the color distribution changes with the index. The colors seen in the reflection images are indeed complementary to those coupled into surface plasmons.[30] Therefore, with a bandpass color filter, the SP resonances would manifest themselves as dark bands in the reflection images. For example, in the color images taken with the red filter at 650 nm (middle column in Fig. 2a), dark bands due to SP resonances with m= -1 are clearly observed for refractive index up to 1.43. For index matching oil ($n_d$ = 1.52), the SP mode with m= -1 is no longer observable and the dark bands from the SP modes with m= +1 and m= -2 emerge. If a green filter centered at 550 nm is used, the m = -1 mode observed for the case of air ($n_d$ = 1.00) rapidly moves out of the range of PDG as the index increases and the dark band due to m = -2 resonance emerges and changes its in-coupling azimuthal angle with increasing refractive index (right column in Fig. 2a). The averaged intensities at each azimuthal angle (angle resolution: 0.2°) were plotted in Fig. 2b. Using our previously developed algorithm[30], the intensity angle distribution can be fitted and the azimuthal angle of the SP resonances for different environmental refractive indexes can be quantitatively determined. Details of the reflection image analysis, fitting procedure and fitting parameters can be found in the Supporting Information. Figure 2b shows the experimentally observed (dots) and fitted (lines) angle distributions at 650 nm and 550 nm. With the fitting, the azimuthal angles of the dark bands (m = -1 resonance at 650 nm) can be quantitatively determined. Since PDG is symmetric about the horizontal line ($\varphi = 0$), each mode results in two dark bands, one in the upper and the other in the lower half of the PDG. This allows us to define our analytical signal as the "open angle" between the two dark bands of the same mode. The open angles obtained from the fitted azimuthal angles of the two peaks are 99.8±0.95°, 252.4±0.37°, 274.6±0.38°, 315.8±0.58° and 346.0±0.38° for surrounding index of 1.00, 1.33, 1.36, 1.39 and 1.43, respectively. Complete data of the evaluated open angles and the corresponding uncertainties can be found in the Supporting Information. It is important to note that the open angle is not a linear function of the index. Therefore, typically used figure of merit, angle changed per RIU, is not suitable for PDG index sensors. In fact, the sensitivity of PDG index sensors is fully design dependent. This makes PDG index sensors unique and advantageous compared to common plasmonic index sensors, such as films or plasmonic nanoparticles. In addition, the spread of the in-coupling angle is also determined by the intrinsic bandwidth of photon-to-plasmon coupling via gratings, which is mainly due to the polarization of the illumination relative to the grating direction, the loss of surface plasmon, and the band-width of illumination and the finite illumination and collection angle.[30] These factors have been included in the analytical model to fit the intensity profiles in Fig. 2b. By plotting the open angle ($2\varphi$) between the two symmetric SP resonance bands (upper panel of Fig. 3a) as a function of the refractive index, calibration curves for quantitative analysis were established. Figure 3b shows the experimental data together with the calibration curves obtained from analytical model for three different grating orders (-2, -1 and +1) at two wavelengths (550 nm and 650 nm). With the wavelengths and grating orders fixed, Eq. (3) describes the relationship between the in-coupling azimuthal angle and the surrounding refractive index. This provides the basis for the analytical prediction of the open angle as a function of the surrounding index. The shaded areas in Fig. 3b mark the ranges of the in-coupling azimuthal angles due to the 40 nm bandwidth of the illumination determined by the bandpass filters. Detailed information about the analytical model can be found in the Supporting Information. The experimental results are in good agreement with the theoretical prediction.

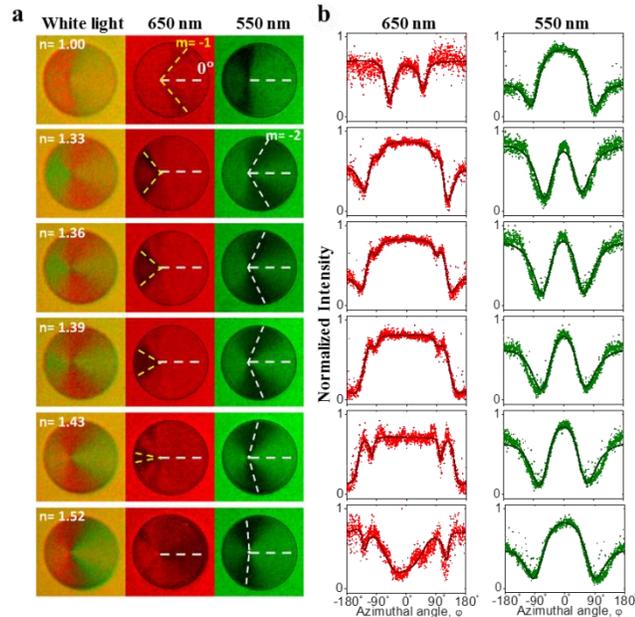

Figure 2. PDG for wide range refractive index sensing. (a) Reflection images of the PDG index sensor illuminated by a nearly normal incident white light without any filter (left column) and with bandpass filters centered at 650 nm (middle column) and 550 nm (right column). Shown from the top to the bottom rows are the reflection images of a PDG covered with air ($n_d$ = 1.00), water ($n_d$ = 1.33), ethanol ($n_d$ = 1.36), tert-butanol ($n_d$ = 1.39), ethylene glycol ($n_d$ = 1.43) and microscope index matching oil ($n_d$ = 1.52). The dashed lines mark the positions of the SP resonance dark bands and the position of the 0° reference. (b) Reflection intensity profiles obtained with the 650 nm (left) and 550 nm (right) bandpass filters. Solid lines are obtained from fitting the experimental data with the analytical model based on Fano resonance model.

Similar to other grating-based sensing platforms, PDGs offer different SP resonances at different colors with different slopes for index sensing. This grants additional flexibility for applications. Since the slope of a calibration curve means sensitivity, one single PDG simultaneously provides various sensitivities for the user to choose. For example, SP modes with small slopes are suitable for applications involving large index change or broad spectral range. SP modes with large slopes are sensitive to small range index sensing. The users can easily choose the SP modes by changing the color of the illumination using bandpass filters. Overall, the PDG reports the variation of the surrounding refractive index as the change in azimuthal angle of the dark bands due to photon-to-plasmon coupling, similar to the pointer in a speedometer and enabling easy and direct evaluation of the surrounding index without spectrometer.

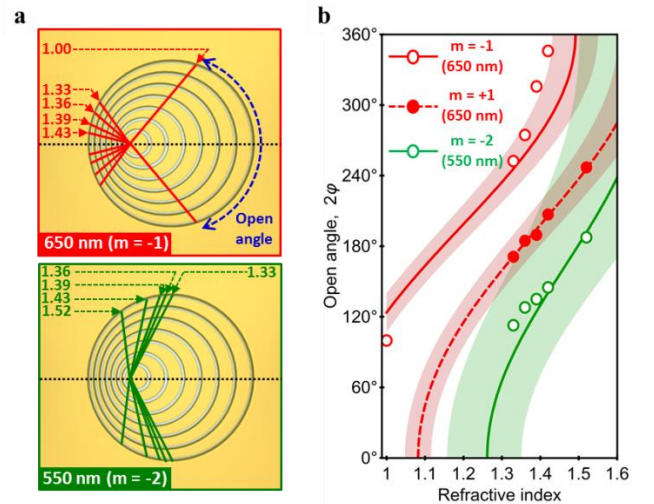

Figure 3. (a) Summary of the in-coupling angle of SP resonance with m = -1 at 650 nm (top panel) and m = -2 at 550 nm (bottom panel) at various environmental indexes. The blue dashed double arrows marks the open angle defined for analytical calibration curves. (b) Calibration curves showing the dependence of the open angle on the surrounding refractive index. Red open dots, red solid dots and green open dots are experimental data of the in-coupling angle of the SP resonance with m = -1 at 650 nm, m = +1 at 650 nm and m = -2 at 550 nm, respectively. Red solid, red dashed and green solid curves are the corresponding results predicted by the analytical model based on Eq. (3). Shaded areas along the curves are the analytically predicted deviation due to the 40 nm bandwidth of the bandpass filter.

Finally, we show that a PDG can be easily designed and optimized for quantitative analysis of a tiny index change. Here, the tiny and nonlinear index change due to the formation of hydrogen bonds in water-ethanol mixture was used as the model system. When water ($n_d$ = 1.3330) and ethanol ($n_d$ = 1.3615) are mixed, the refractive index of the mixture solution can go up to $n_d$ = 1.3650, exceeding that of pure ethanol. This is due to the intermolecular interaction between water and ethanol molecules. The nonlinear refractive index change has been broadly used to evaluate the density of the water-ethanol mixture and to indicate their mixing ratio.[46] The peculiar enhancement of the refractive index is very small and is difficult to be directly observed by standard plasmonic index sensing platforms. By using specially designed PDG with high sensitivity for tiny index change, it is possible to resolve and follow the index change of the mixture of ethanol and water between 1.3330 and 1.3650 and clearly observe the peculiar index enhancement.

To resolve the small index change as a function of molar ratio of ethanol ($\chi_{EtOH}$), the PDG index sensor has been designed to have a radius increment $\Delta r$ = 390 nm and a displacement of center $d$ = 40 nm. The choice of these parameters allows the PDG index sensor to have best sensitivity within the range of index variation between 1.3330 and 1.3650. The analytical performance of a PDG with a slightly different design ($\Delta r$ = 390 nm, $d$ = 20 nm) is also provided in the Supporting Information for reference. To enhance the sensitivity, a laser line filter centered at 632.8 nm with a bandwidth of 3 nm is used to produce narrow band illumination. Figure 4a shows the raw reflection images of the same PDG in contact with a series of water-ethanol mixture. Dark bands are clearly observed at different azimuthal angles depending on the molar ratio of ethanol ($\chi_{EtOH}$). By fitting the intensity angle distribution profiles (Fig. 4b), the azimuthal angle of SP resonances and the index of each mixture solution can be precisely quantified. The indexes obtained from PDG are plotted in Fig. 4c together with experimental data from previous works using refractometer-based methods. The agreement is very good and the peculiar index enhancement due to hydrogen bonds around $\chi_{EtOH}$=0.5, i.e. the index of the mixture exceeds that of pure ethanol, is also clearly observed.[46-48] Table 1 summarizes the obtained indexes with errors obtained in this work and compares it with the data from Ref. 47. This example demonstrates PDG's ability to "zoom in" to a small index window to achieve ultimate resolution in small refractive index window.

anol. Solid traces are obtained from fitting the experimental data with an analytical model based on Fano resonance model.[30] (c) The index of the water-ethanol mixture as a function of the molar ratio of ethanol $\chi_{EtOH}$. Red open stars mark the data points obtained with the PDG index sensor in this work. Green crosses, grey open triangles and the violet open squares mark the data from Refs. 46, 47 and 48, respectively. The SEM image of the PDG used here is shown in the insert of (c). The scale bar is 2.5 µm.

Table 1. Summary of the open angles ($2\varphi$) and the corresponding refractive indexes ($n_d$) of the water-ethanol mixture solutions. Data from Ref. 47 are also listed for comparison.

| $\chi_{EtOH}$ | $2\varphi$ | $n_d$ | Ref. 47 |
|---|---|---|---|
| 0 | 133.2±0.51 | 1.3345±0.00102 | 1.3330 |
| 0.05 | 141.4±0.60 | 1.3424±0.00117 | - |
| 0.10 | 143.2±0.46 | 1.3442±0.00090 | 1.3457 |
| 0.20 | 150.8±0.44 | 1.3517±0.00089 | 1.3535 |
| 0.50 | 160.4±0.39 | 1.3615±0.00079 | 1.3604 |
| 1.00 | 156.4±0.36 | 1.3574±0.00073 | 1.3593 |

**Conclusion** We have demonstrated the design and applications of PDG index sensors for large range and small range of index change. Since the working principle of PDG index sensor is based on photon-to-plasmon coupling via azimuthal angle dependent plasmonic gratings, the sensitivity of a PDG index sensor can be easily optimized for specific applications by choosing suitable design parameters. With an optimal design, we have demonstrated the ability of PDG to quantitatively measure very small index change of water-ethanol mixture. The peculiar index enhancement in water-ethanol solution has been clearly observed and precisely quantified. Since PDG is a planar and dispersive microstructure, it can be easily integrated into microfluidic channels for on-site spectrometer-free index sensing. With surface functionalization, PDGs can also perform spectrometer-free biochemical analysis. By incorporating active controllable materials, it is also possible to actively control the optical response of a PDG. We anticipate various sensing applications using PDGs.

### Supporting Information

Supporting Information Available: Reflection image analysis, reflection intensity profile fitting, uncertainty of peak position, analytical model for calibration curves, Small-range index sensing data from a PDG with d= 20 nm and $\Delta r$ = 390 nm

### Corresponding Author

* jer-shing.huang@leibniz-ipht.de

### Author Contributions

F.-C.L., K.-M.S. and J.-S.H. conceived the idea. F.-C.L., K.-M.S. and Y.-X.H. fabricated the structures. F.-C.L. and K.-M.S. performed the optical measurement and numerical simulations. All coauthors contributed to the data analysis, interpretation and manuscript preparation. J.P. and J.-S.H. supervised the research. # These authors contributed equally.

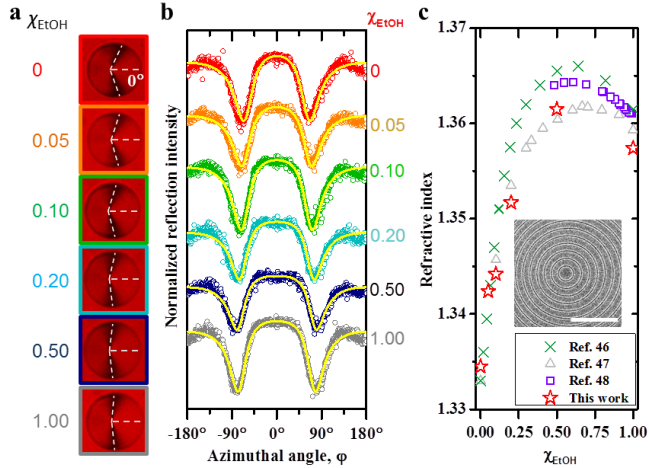

Figure 4. (a) Reflection images of the PDG index sensor illuminated by a near-normally incident white light filtered by a narrow band filter centered at 632.8 nm. Shown from the top to the bottom rows are the reflection images of the PDG index sensor covered with water-ethanol mixture with molar ratio of ethanol $\chi_{EtOH}$ = 0, 0.05, 0.10, 0.20, 0.50 and 1.00. The dashed lines mark the azimuthal angles of the SP resonant bands with m = -1 and the position of the 0° reference. (b) Intensity angle distribution extracted from the reflection images in (a) at corresponding molar ratio of eth-

## ACKNOWLEDGMENT

Financial support by the Thuringian State Government within its ProExcellence initiative (APC²⁰²⁰), the DFG (HU2626/3-1 and CRC 1375 NOA) and the Ministry of Science and Technology of Taiwan under contract no. MOST-103-2113-M-007-004-MY3 are acknowledged

## REFERENCES

1. Sepúlveda, B.; Angelomé, P. C.; Lechuga, L. M.; Liz-Marzán, L. M., LSPR-based nanobiosensors. *Nano Today* **2009,** *4* (3), 244-251.
2. Dreaden, E. C.; Alkilany, A. M.; Huang, X.; Murphy, C. J.; El-Sayed, M. A., The golden age: gold nanoparticles for biomedicine. *Chemical Society Reviews* **2012,** *41* (7), 2740-2779.
3. Polo, E.; Poupard, M. F. N.; Guerrini, L.; Taboada, P.; Pelaz, B.; Alvarez-Puebla, R. A.; del Pino, P., Colloidal bioplasmonics. *Nano Today* **2018,** *20*, 58-73.
4. Kailasa, S. K.; Koduru, J. R.; Desai, M. L.; Park, T. J.; Singhal, R. K.; Basu, H., Recent progress on surface chemistry of plasmonic metal nanoparticles for colorimetric assay of drugs in pharmaceutical and biological samples. *TrAC Trends in Analytical Chemistry* **2018,** *105*, 106-120.
5. Cortie, M. B.; McDonagh, A. M., Synthesis and optical properties of hybrid and alloy plasmonic nanoparticles. *Chemical Reviews* **2011,** *111* (6), 3713-3735.
6. Olson, J.; Dominguez-Medina, S.; Hoggard, A.; Wang, L.-Y.; Chang, W.-S.; Link, S., Optical characterization of single plasmonic nanoparticles. *Chemical Society Reviews* **2015,** *44* (1), 40-57.
7. Chen, Y.-Y.; Chang, H.-T.; Shiang, Y.-C.; Hung, Y.-L.; Chiang, C.-K.; Huang, C.-C., Colorimetric assay for lead ions based on the leaching of gold nanoparticles. *Analytical Chemistry* **2009,** *81* (22), 9433-9439.
8. Zhang, Z.; Chen, Z.; Qu, C.; Chen, L., Highly sensitive visual detection of copper ions based on the shape-dependent LSPR spectroscopy of gold nanorods. *Langmuir* **2014,** *30* (12), 3625-3630.
9. Wadell, C.; Syrenova, S.; Langhammer, C., Plasmonic hydrogen sensing with nanostructured metal hydrides. *ACS Nano* **2014,** *8* (12), 11925-11940.
10. Ng, K. C.; Lin, F.-C.; Yang, P.-W.; Chuang, Y.-C.; Chang, C.-K.; Yeh, A.-H.; Kuo, C.-S.; Kao, C.-R.; Liu, C.-C.; Jeng, U. S.; Huang, J.-S.; Kuo, C.-H., Fabrication of Bimetallic Au–Pd–Au Nanobricks as an Archetype of Robust Nanoplasmonic Sensors. *Chemistry of Materials* **2018,** *30* (1), 204-213.
11. Zhang, B.; Kumar, R. B.; Dai, H.; Feldman, B. J., A plasmonic chip for biomarker discovery and diagnosis of type 1 diabetes. *Nature Medicine* **2014,** *20* (8), 948.
12. Park, J.; Im, H.; Hong, S.; Castro, C. M.; Weissleder, R.; Lee, H., Analyses of Intravesicular Exosomal Proteins Using a Nano-Plasmonic System. *ACS Photonics* **2017,** *5* (2), 487-494.
13. Tittl, A.; Yin, X.; Giessen, H.; Tian, X.-D.; Tian, Z.-Q.; Kremers, C.; Chigrin, D. N.; Liu, N., Plasmonic Smart Dust for Probing Local Chemical Reactions. *Nano Letters* **2013,** *13* (4), 1816-1821.
14. Rodal-Cedeira, S.; Montes-García, V.; Polavarapu, L.; Solís, D. M.; Heidari, H.; La Porta, A.; Angiola, M.; Martucci, A.; Taboada, J. M.; Obelleiro, F., Plasmonic au@ pd nanorods with boosted refractive index susceptibility and sers efficiency: A multifunctional platform for hydrogen sensing and monitoring of catalytic reactions. *Chemistry of Materials* **2016,** *28* (24), 9169-9180.
15. Shen, Y.; Zhou, J.; Liu, T.; Tao, Y.; Jiang, R.; Liu, M.; Xiao, G.; Zhu, J.; Zhou, Z.-K.; Wang, X., Plasmonic gold mushroom arrays with refractive index sensing figures of merit approaching the theoretical limit. *Nature Communications* **2013,** *4*, 2381.
16. Raza, S.; Toscano, G.; Jauho, A.-P.; Mortensen, N. A.; Wubs, M., Refractive-index sensing with ultrathin plasmonic nanotubes. *Plasmonics* **2013,** *8* (2), 193-199.
17. Wang, F.; Shen, Y. R., General properties of local plasmons in metal nanostructures. *Physical Review Letters* **2006,** *97* (20), 206806.
18. Liu, C.-Y.; Tseng, W.-L., Colorimetric assay for cyanide and cyanogenic glycoside using polysorbate 40-stabilized gold nanoparticles. *Chemical Communications* **2011,** *47* (9), 2550-2552.
19. Zhou, Y.; Dong, H.; Liu, L.; Xu, M., Simple Colorimetric Detection of Amyloid β‐peptide (1–40) based on Aggregation of Gold Nanoparticles in the Presence of Copper Ions. *Small* **2015,** *11* (18), 2144-2149.
20. Mesch, M.; Weiss, T.; Schäferling, M.; Hentschel, M.; Hegde, R. S.; Giessen, H., Highly Sensitive Refractive Index Sensors with Plasmonic Nanoantennas–Utilization of Optimal Spectral Detuning of Fano Resonances. *ACS Sensors* **2018,** *3* (5), 960-966.
21. Zhang, Y.; Liu, W.; Li, Z.; Li, Z.; Cheng, H.; Chen, S.; Tian, J., High-quality-factor multiple Fano resonances for refractive index sensing. *Optics Letters* **2018,** *43* (8), 1842-1845.
22. King, N. S.; Liu, L.; Yang, X.; Cerjan, B.; Everitt, H. O.; Nordlander, P.; Halas, N. J., Fano resonant aluminum nanoclusters for plasmonic colorimetric sensing. *ACS Nano* **2015,** *9* (11), 10628-10636.
23. Liu, D.; Fang, L.; Zhou, F.; Li, H.; Zhang, T.; Li, C.; Cai, W.; Deng, Z.; Li, L.; Li, Y., Ultrasensitive and Stable Au Dimer‐Based Colorimetric Sensors Using the Dynamically Tunable Gap‐Dependent Plasmonic Coupling Optical Properties. *Advanced Functional Materials* **2018,** *28* (18), 1707392.
24. Fan, J. R.; Zhu, J.; Wu, W. G.; Huang, Y., Plasmonic Metasurfaces Based on Nanopin‐Cavity Resonator for Quantitative Colorimetric Ricin Sensing. *Small* **2017,** *13* (1), 1601710.
25. Maity, D.; Bhatt, M.; Paul, P., Calix [4] arene functionalized gold nanoparticles for colorimetric and bare-eye detection of iodide in aqueous media and periodate aided enhancement in sensitivity. *Microchimica*

*Acta* **2015,** *182* (1-2), 377-384.
26. Kretschmann, E.; Raether, H., Notizen: Radiative Decay of Non Radiative Surface Plasmons Excited by Light. In *Zeitschrift für Naturforschung A*, 1968; Vol. 23, p 2135.
27. Otto, A., Excitation of Nonradiative Surface Plasma Waves in Silver by the Method of Frustrated Total Reflection. *Z Phys* **1968,** *216*, 398-410.
28. Renger, J.; Quidant, R.; van Hulst, N.; Palomba, S.; Novotny, L., Free-Space Excitation of Propagating Surface Plasmon Polaritons by Nonlinear Four-Wave Mixing. *Phys Rev Lett* **2009,** *103* (26), 266802.
29. Li, G.; Zhang, S.; Zentgraf, T., Nonlinear photonic metasurfaces. *Nature Reviews Materials* **2017,** *2* (5), 17010.
30. See, K.-M.; Lin, F.-C.; Huang, J.-S., Design and characterization of a plasmonic Doppler grating for azimuthal angle-resolved surface plasmon resonances. *Nanoscale* **2017,** *9* (30), 10811-10819.
31. Lee, K.-L.; Chen, P.-W.; Wu, S.-H.; Huang, J.-B.; Yang, S.-Y.; Wei, P.-K., Enhancing surface plasmon detection using template-stripped gold nanoslit arrays on plastic films. *ACS Nano* **2012,** *6* (4), 2931-2939.
32. López-Tejeira, F.; Rodrigo, S. G.; Martín-Moreno, L.; García-Vidal, F. J.; Devaux, E.; Ebbesen, T. W.; Krenn, J. R.; Radko, I.; Bozhevolnyi, S. I.; González, M. U., Efficient unidirectional nanoslit couplers for surface plasmons. *Nature Physics* **2007,** *3* (5), 324.
33. Genevet, P.; Capasso, F., Holographic optical metasurfaces: a review of current progress. *Reports on Progress in Physics* **2015,** *78* (2), 024401.
34. Sobhani, A.; Knight, M. W.; Wang, Y.; Zheng, B.; King, N. S.; Brown, L. V.; Fang, Z.; Nordlander, P.; Halas, N. J., Narrowband photodetection in the near-infrared with a plasmon-induced hot electron device. *Nature Communications* **2013,** *4*, 1643.
35. Quaranta, G.; Basset, G.; Benes, Z.; Martin, O. J.; Gallinet, B., Light refocusing with up-scalable resonant waveguide gratings in confocal prolate spheroid arrangements. *Journal of Nanophotonics* **2018,** *12* (1), 016004.
36. Quaranta, G.; Basset, G.; Martin, O. J.; Gallinet, B., Recent advances in resonant waveguide gratings. *Laser & Photonics Reviews* **2018,** *12* (9), 1800017.
37. Quaranta, G.; Basset, G.; Martin, O. J.; Gallinet, B., Color-selective and versatile light steering with up-scalable subwavelength planar optics. *ACS Photonics* **2017,** *4* (5), 1060-1066.
38. Li, R.; Wu, D.; Liu, Y.; Yu, L.; Yu, Z.; Ye, H., Infrared plasmonic refractive index sensor with ultra-high figure of merit based on the optimized all-metal grating. *Nanoscale Research Letters* **2017,** *12* (1), 1.
39. Dai, Y.; Xu, H.; Wang, H.; Lu, Y.; Wang, P., Experimental demonstration of high sensitivity for silver rectangular grating-coupled surface plasmon resonance (SPR) sensing. *Optics Communications* **2018,** *416*, 66-70.
40. Bijalwan, A.; Rastogi, V., Design Analysis of Refractive Index Sensor with High Quality Factor Using Au-Al 2 O 3 Grating on Aluminum. *Plasmonics* **2018**, 1-6.
41. Bdour, Y.; Escobedo, C.; Sabat, R. G., Wavelength-selective plasmonic sensor based on chirped-pitch crossed surface relief gratings. *Optics Express* **2019,** *27* (6), 8429-8439.
42. Nair, S.; Escobedo, C.; Sabat, R. G., Crossed surface relief gratings as nanoplasmonic biosensors. *ACS Sensors* **2017,** *2* (3), 379-385.
43. Triggs, G. J.; Wang, Y.; Reardon, C. P.; Fischer, M.; Evans, G. J.; Krauss, T. F., Chirped guided-mode resonance biosensor. *Optica* **2017,** *4* (2), 229-234.
44. Huang, J.-S.; Callegari, V.; Geisler, P.; Brüning, C.; Kern, J.; Prangsma, J. C.; Wu, X.; Feichtner, T.; Ziegler, J.; Weinmann, P., Atomically flat single-crystalline gold nanostructures for plasmonic nanocircuitry. *Nature Communications* **2010,** *1*, 150.
45. Dai, W.-H.; Lin, F.-C.; Huang, C.-B.; Huang, J.-S., Mode conversion in high-definition plasmonic optical nanocircuits. *Nano letters* **2014,** *14* (7), 3881-3886.
46. Koohyar, F.; Kiani, F.; Sharifi, S.; Sharifirad, M.; Rahmanpour, S. H., Study on the change of refractive index on mixing, excess molar volume and viscosity deviation for aqueous solution of methanol, ethanol, ethylene glycol, 1-propanol and 1, 2, 3-propantriol at T= 292.15 K and atmospheric pressure. *Research Journal of Applied Sciences, Engineering and Technology* **2012,** *4* (17), 3095-3101.
47. Jose V. H., R. B., Refractive Indices, Densities and Excess Molar Volumes of Monoalcohols + Water *Journal of Solution Chemistry* **2006,** *35*, 1315-1328.
48. Jiménez Riobóo, R. J., Philipp, M., Ramos, M. A., Krüger, J. K. , Concentration and temperature dependence of the refractive index of ethanol-water mixtures: Influence of intermolecular interactions. *The European Physical Journal E* **2009,** *30*, 19-26.

For the Table of Content Only

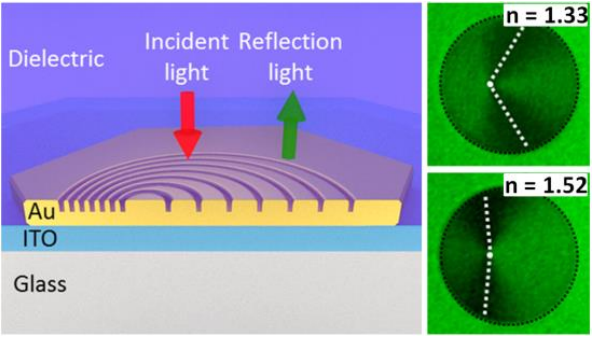

# *Supporting Information*

## Designable spectrometer-free index sensing using plasmonic Doppler gratings

Fan-Cheng Lin,[†,#] Kel-Meng See,[†,#] Lei Ouyang,[‡,§] You-Xin Huang,[†] Yi-Ju Chen,[‡] Jürgen Popp[‡,§] and Jer-Shing Huang[*,†,‡,⊥,∥]

[†]Department of Chemistry, National Tsing Hua University, Hsinchu 30013, Taiwan

[‡]Leibniz Institute of Photonic Technology, Albert-Einstein Straße 9, D-07745 Jena, Germany

[§]Institute of Physical Chemistry and Abbe Center of Photonics, Friedrich-Schiller-Universität Jena, Helmholtzweg 4, D-07743 Jena, Germany

[⊥]Research Center for Applied Sciences, Academia Sinica, 128 Sec. 2, Academia Road, Nankang District, Taipei 11529, Taiwan

[∥]Department of Electrophysics, National Chiao Tung University, Hsinchu 30010, Taiwan

* jer-shing.huang@leibniz-ipht.de

Content of the Supporting Information:

# Reflection Image Analysis

The reflection images are analyzed to get the intensity distribution at different azimuthal angles. These images are first spatially filtered by a mask to remove the pixels in the irrelevant region out of the structured area. The spatially filtered color images are then converted into gray scale intensity image using the standard default color space,

$$I=0.2989 * R + 0.5870 * G + 0.1140 * B \quad [S1]$$

The code used to analyze the reflection images in MATLAB is as follow:

```
color=imread('PDG-Air-650.jpg');     % load figure
gray=rgb2gray(color);% return the RGB color to gray scale
cgray=imcomplement(gray);
imtool(cgray); % change the histogram range
% adjust ' contrast'
% Export to workspace. Set name ('ncgray')
n=size(ncgray);
n1=n(1,1);
n2=n(1,2);
matrix=ones(n1*n2,4);
for ny=1:n1
for nx=1:n2
matrix(n2*(ny-1) +nx,1)= n2*(ny-1) +nx;
matrix(n2*(ny-1) +nx,2)=nx;
matrix(n2*(ny-1) +nx,3)=ny;
matrix(n2*(ny-1) +nx,4)=ncgray(ny,nx);
end
end
dlmwrite('PDG-Air-650.txt',matrix);
```

The azimuthal angle for each pixel is determined based on the formula

$$\varphi[x_a, y_b] = \tan^{-1}\left(\frac{y_b+y_0}{x_a-x_0}\right) \quad [S2]$$

, where $x_a$ and $y_b$ are the pixel coordinates of the matrix, while $x_0$ and $y_0$ are the pixel coordinate at the position of the smallest ring with zero diameter (as shown below)

S-2

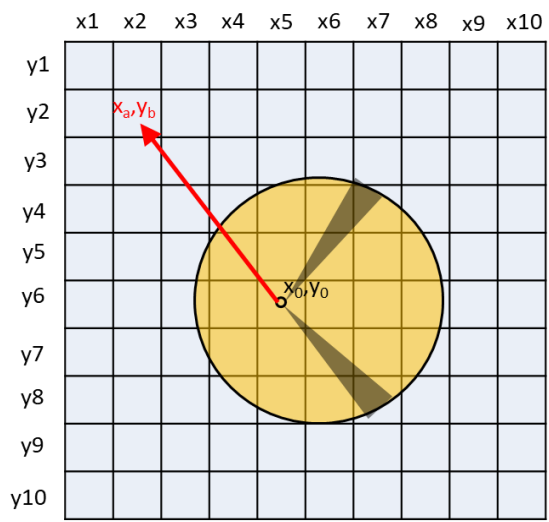

*Relative angle of pixel respect to PDG center*

$$\varphi = \tan^{-1}\frac{y_b - y_0}{x_a - x_0}$$

**Figure S1.** Schematic illustration of pixel to azimuthal angle conversion. For the clarity of illustration, the size of pixels has been enlarged. Therefore, the pixel size relative to the area of PDG is not realistic. The actual imaging area of a pixel is estimated to be around 207*207 nm$^2$.

Rounding the azimuthal angles to the nearest integer of 0.2 degree. Intensities of all pixels within the bandwidth of 0.2 degree were summed up and divided by the pixel number to obtain the averaged intensity. The rounding was done by applying the MROUND function in Excel (Microsoft) and the averaged intensity was obtained using AVERAGEIF function in Excel. The obtained intensity profiles are then normalized to the maximum pixel value. The normalized intensity is then plotted as a function of the azimuthal angle to demonstrate the intensity angle distribution profile.

**Table S1.** Examples for the rounding procedure and the averaged intensity.

| Original angle: tan$^{-1}$ [(y$_b$-y$_0$)/ (x$_a$-x$_0$)] | Rounded angle: bandwidth: 0.2° | Original intensity | Average intensity after processed |
|---|---|---|---|
| 46.1233 | 46.2 | 0.6 | 0.623685 |
| 46.1691 | | 0.6 | |
| 46.2189 | | 0.64737 | |
| 46.2730 | | 0.64737 | |
| 46.3322 | 46.4 | 0.64737 | 0.66579 |
| 46.3972 | | 0.6 | |
| 46.3972 | | 0.6 | |
| 46.4321 | | 0.9 | |
| 46.4688 | | 0.6 | |
| 46.4688 | | 0.64737 | |

# Reflection Intensity Profile Fitting

The azimuthal angle of the dark band is quantified by fitting the angle distribution intensity profiles with the following equation,

$$I(\varphi) = \sum_m \left( \frac{\left(\frac{P(\varphi+x_0)^2 - P(\varphi_m)^2}{2 w_m P_m} + q_m\right)^2 + b_m}{\left(\frac{P(\varphi+x_0)^2 - P(\varphi_m)^2}{2 w_m P_m}\right)^2 + 1} - 1 \right) \cdot \sqrt{(\cos(\varphi + x_0 - \gamma))^2 + (1 - \eta)(\sin(\varphi + x_0 - \gamma))^2} \cdot A_m + y_0 \quad [S3]$$

$I(\varphi)$: reflection intensity at azimuthal angle $\varphi$, m: resonance order; $P(\varphi)$: azimuthal angle-dependent periodicity, $P(\varphi_m)$: resonant grating periodicity for the $m^{th}$-order resonance; $w_m$: spectral width of the Fano resonance for the $m^{th}$-order resonance; $q_m$: asymmetry parameter for the $m^{th}$-order resonance; $b_m$: modulation damping parameter for the $m^{th}$-order resonance; $\gamma$: angle between incident light polarization and the edge of the rings; $\eta$: degree of polarization of the excitation in the experiment; $A_m$: amplitude of the $m^{th}$-order resonance; $y_0$: amplitude offset; $x_0$: angle offset;

This fitting formula has been developed in our previous work [ref. 30] to fit the intensity profile of reflection and transmission. This fitting formula is based on Fano-like resonance model and takes into account the broadening due to the loss of SPs, the coupling to dark modes, the degree of polarization of the illumination and the incident polarization angle with respect to the PDG grating edge. In our previous work [ref. 30], we have demonstrated PDG-based color sorter, where the coupling between the grating resonance (narrow band) and the transmission of the slit (broad band) has led to Fano-like resonance revealed as very asymmetric transmission intensity profiles along azimuthal angle (Fig. 3d in ref. 30). In this work, the PDG index sensors contain, however, only grooves without any slit. Therefore, there is no Fano-like resonance and the reflection intensity profiles are very symmetric, as expected for typical plasmonic grating couplers. As a result, when applying Eq. S3 to fit the intensity profile, we have obtained very small $q_m$ (see Table S2 below). This is equivalent to fitting the profiles with a Cauchy–Lorentz distribution. The key point here is to determine the peak position and its uncertainty (i.e. the azimuthal angle of the dark bands) so that the surrounding index can be calculated using Eq. 3 in the main text. The fitting parameters used for the experimental data in Fig. 2b and Fig. 4b in the main text are summarized in Table S2 below. The subscripts denote the order of the grating resonance.

**Table S2.** Parameters used for fitting reflection intensity profiles for large range index sensing (Fig. 2b) and for small range index sensing (Fig. 4b).

| Fitting parameters used for the profiles in the **left** column of Fig. 2b (**650 nm**) | | | | | | | |
|---|---|---|---|---|---|---|---|
| parameter | | $n = 1.00$ | $n = 1.33$ | $n = 1.36$ | $n = 1.39$ | $n = 1.43$ | $n = 1.52$ |
| $P_m$ | $m = -1$ | 578.38 | 408.19 | 397.61 | 376.13 | 372.22 | - |
| | $m = +1$ | - | 495.15 | 470.73 | 471.66 | 451.01 | 402.28 |
| | $m = -2$ | - | - | - | - | - | 631.25 |
| $w_m$ | $m = -1$ | 33.34 | 35.60 | 38.24 | 40.89 | 21.14 | - |
| | $m = +1$ | - | 30.59 | 27.02 | 29.27 | 17.36 | 14.90 |
| | $m = -2$ | - | - | - | - | - | 163.08 |
| $A_m$ | $m = -1$ | 1.00 | 1.00 | 1.00 | 1.00 | 1.00 | - |
| | $m = +1$ | - | 0.27 | 0.35 | 0.14 | 0.49 | 1.40 |
| | $m = -2$ | - | - | - | - | - | 1.00 |
| $q_m$ | $m = -1$ | 0.045 | -0.005 | 0.095 | 0.145 | 0.215 | - |
| | $m = +1$ | - | -0.003 | -0.250 | 0.00 | 0.101 | -0.111 |
| | $m = -2$ | - | - | - | - | - | 0.093 |
| $b_m$ | $m = -1$ | 0.3684 | 0.0710 | 0.1617 | 0.3134 | 0.5154 | - |
| | $m = +1$ | - | 0.0358 | 0.0017 | 0.00 | 0.0074 | 0.7784 |
| | $m = -2$ | - | - | - | - | - | 0.1856 |
| $\Delta r$ (nm) | | 500 | 500 | 500 | 500 | 500 | 500 |
| $d$ (nm) | | 140 | 140 | 140 | 140 | 140 | 140 |
| $\eta$ | | 0.50 | 0.46 | 0.45 | 0.02 | 0.50 | 0.40 |
| $\gamma$ | | 163.84 | 172.22 | 178.02 | 200.01 | 171.06 | 168.28 |
| $y_0$ | | 0.73 | 0.88 | 0.84 | 0.79 | 0.68 | 1.02 |

| Fitting parameters used for the profiles in the **right** column of Fig. 2b (**550 nm**) | | | | | | | |
|---|---|---|---|---|---|---|---|
| parameter | | $n = 1.00$ | $n = 1.33$ | $n = 1.36$ | $n = 1.39$ | $n = 1.43$ | $n = 1.52$ |
| $P_m$ | $m = -1$ | 475.72 | - | - | - | - | - |
| | $m = -2$ | - | 584.17 | 563.70 | 564.19 | 546.35 | 476.72 |
| $w_m$ | $m = -1$ | 72.81 | - | - | - | - | - |
| | $m = -2$ | - | 85.73 | 90.01 | 103.99 | 99.62 | 106.32 |
| $A_m$ | $m = -1$ | 1.00 | - | - | - | - | - |
| | $m = -2$ | - | 1.00 | 1.00 | 1.00 | 1.00 | 1.00 |
| $q_m$ | $m = -1$ | 0.144 | - | - | - | - | - |
| | $m = -2$ | | 0.240 | 0.188 | 0.268 | 0.218 | 0.043 |
| $b_m$ | $m = -1$ | 0.00 | - | - | - | - | - |
| | $m = -2$ | - | 0.00 | 0.00 | 0.00 | 0.00 | 0.00 |
| $\Delta r$ | | 500 | 500 | 500 | 500 | 500 | 500 |
| $d$ | | 140 | 140 | 140 | 140 | 140 | 140 |
| $\eta$ | | 0.275 | 0.084 | 0.00 | 0.059 | 0.049 | 0.105 |
| $\gamma$ | | 180 | 150 | 150 | 150 | 150 | 150 |
| $y_0$ | | 0.86 | 1.15 | 1.15 | 1.10 | 1.09 | 1.02 |

| *Fitting parameters used for the profiles in Fig. 4b* | | | | | | | |
|---|---|---|---|---|---|---|---|
| *parameter* | | $\chi_{EtOH} = 0$ | $\chi_{EtOH} = 0.05$ | $\chi_{EtOH} = 0.10$ | $\chi_{EtOH} = 0.20$ | $\chi_{EtOH} = 0.50$ | $\chi_{EtOH} = 1.00$ |
| $P_m$ | | 407.1786 | 404.6105 | 403.1063 | 401.2025 | 397.7266 | 398.6259 |
| $w_m$ | | 14.544 | 14.531 | 14.009 | 13.864 | 14.032 | 13.929 |
| $A_m$ | $m = -1$ | 1 | 1 | 1 | 1 | 1 | 1 |
| $q_m$ | | 0.16762 | 0.15092 | 0.14461 | 0.14547 | 0.12751 | 0.1382 |
| $b_m$ | | 0.20754 | 0.30932 | 0.15914 | 0.35624 | 0.40636 | 0.17801 |
| $\Delta r$ | | 390 | 390 | 390 | 390 | 390 | 390 |
| $d$ | | 40 | 40 | 40 | 40 | 40 | 40 |
| $\eta$ | | 0 | 0 | 0 | 0 | 0 | 0 |
| $\gamma$ | | 0 | 0 | 0 | 0 | 0 | 0 |
| $y_0$ | | 0.96994 | 0.956 | 0.97557 | 0.82355 | 0.80369 | 0.91962 |

# Uncertainty of Peak Position

The uncertainty in peak position ($x - \bar{x}_m$) was estimated using the following equation (*Rev. Sci. Instrum.*, **1986**, *57*, 1152-1157)

$$x - \bar{x}_m = \Gamma \frac{1}{\sqrt{S}} \sqrt{\frac{\Delta}{N_\Gamma t F_p(t)}} \quad [S4]$$

, where the dimensionless quantity *t* is the amplitude threshold level. $\Gamma$ is the considered data range under the threshold *t*. The choice of $t = 2$ corresponds to using the measurement signal falling within ± 2$\Gamma$ about the peak maximum. $\sqrt{S}$ reflects the signal to noise ratio for Poisson statistics. $\Delta$ is a numerical constant that depends on the chosen confidence level *c*. For $c = 0.68$ ("±1 σ"), $\Delta$ takes values of 1, 2.3, or 3.5 for opening one, two or three parameters for fitting. In this work, three parameters have been open for fitting. Therefore, $\Delta = 3.5$ has been used. $N_\Gamma$ is the number of data points within the distance $\Gamma$ from the peak center. The dependence on the threshold setting for Poisson-distributed noise is given by the function $\sqrt{t F_p(t)}$. The value of $\sqrt{t F_p(t)}$ is obtained from the reference (*Rev. Sci. Instrum.*, **1986**, *57*, 1152-1157). In our work, it is 1.3 for data with Poisson noise at t = 2. Taking the intensity angle distribution of a PDG in air illuminated with red light (λ=650 nm) as an example, SNR was calculated to be 10.54, the uncertainty in peak position is thus

$$x - \bar{x}_m = 92° \times \frac{1}{10.54} \sqrt{\frac{3.5}{461 \times 1.3}} = 0.67°$$

Since our analytical signal "open angle" is the difference between the peak positions of the two peaks of one mode in an intensity profile, the uncertainty of the open angle can be obtained by pooling the uncertainties of the two peaks, i.e. $\Delta 2\varphi = \pm((\Delta\varphi_a)^2 + (\Delta\varphi_b)^2)^{1/2}$. The calculated uncertainties for all fitted peak positions and the pooled uncertainties of all open angles are summarized in Table S3.

**Table S3.** The calculated uncertainties for all fitted peak positions in Figs. 2b and 4b.

| *Uncertainties of all peaks in the left column of Fig. 2b (m = -1, 650 nm)* | | | |
|---|---|---|---|
| surrounding index | left peak: $\varphi_a \pm \Delta\varphi_a$ | right peak: $\varphi_b \pm \Delta\varphi_b$ | open angle ($2\varphi$): $\varphi_b - \varphi_a \pm ((\Delta\varphi_a)^2 + (\Delta\varphi_b)^2)^{1/2}$ |
| 1.00 | -50.2°±0.67° | 49.6°± 0.68° | 99.8°± 0.95° |
| 1.33 | -126.4°± 0.27° | 126.0°±0.25° | 252.4°±0.37° |
| 1.36 | -137.4°±0.27° | 137.2°±0.27° | 274.6°±0.38° |
| 1.39 | -157.6°±0.38° | 158.2°±0.44° | 315.8°± 0.58° |
| 1.43 | -180.0°±0.23° | 166.0°±0.30° | 346.0°± 0.38° |
| 1.52 | - | - | - |

| *Uncertainties of all peaks in the left column of Fig. 2b (m = +1, 650 nm)* | | | |
|---|---|---|---|
| surrounding index | left peak: $\varphi_a \pm \Delta\varphi_a$ | right peak: $\varphi_b \pm \Delta\varphi_b$ | open angle ($2\varphi$): $\varphi_b - \varphi_a \pm ((\Delta\varphi_a)^2 + (\Delta\varphi_b)^2)^{1/2}$ |
| 1.00 | - | - | - |
| 1.33 | -85.0°±0.49° | 86.8°±0.43° | 172.4°±0.62° |
| 1.36 | -92.2°±0.40° | 92.4°±0.42° | 184.6°±0.58° |
| 1.39 | -94.8°±0.52° | 94.8°±0.59° | 189.6°±0.79° |
| 1.43 | -103.2°±0.47° | 103.4°±0.52° | 206.6°±0.70° |
| 1.52 | -124.0°±1.40° | 123.8°±0.98° | 247.8°±1.71° |

| *Uncertainties of all peaks in the right column of Fig. 2b (m=-2, 550 nm)* | | | |
|---|---|---|---|
| surrounding index | left peak: $\varphi_a \pm \Delta\varphi_a$ | right peak: $\varphi_b \pm \Delta\varphi_b$ | open angle ($2\varphi$): $\varphi_b - \varphi_a \pm ((\Delta\varphi_a)^2 + (\Delta\varphi_b)^2)^{1/2}$ |
| 1.00 | - | - | - |
| 1.33 | -56.4°± 0.56° | 56.4°± 0.61° | 112.8°±0.83° |
| 1.36 | -63.4°± 0.42° | 63.4°±0.45° | 126.8°±0.62° |
| 1.39 | -66.8°± 0.35° | 68.2°±0.39° | 135.0°±0.53° |
| 1.43 | -71.6°± 0.48° | 72.8°±0.45° | 144.4°±0.66° |
| 1.52 | -93.0°± 0.45° | 94.4°±0.44° | 187.4°±0.62° |

| *Uncertainties of all peaks in Fig. 4b* | | | |
|---|---|---|---|
| $\chi_{EtOH}$ | left peak: $\varphi_a \pm \Delta\varphi_a$ | right peak: $\varphi_b \pm \Delta\varphi_b$ | open angle ($2\varphi$): $\varphi_b - \varphi_a \pm ((\Delta\varphi_a)^2 + (\Delta\varphi_b)^2)^{1/2}$ |
| 0 | -66.6°±0.35° | 66.6°± 0.37° | 133.2°±0.51° |
| 0.05 | -70.8°±0.43° | 70.6°±0.41° | 141.4°±0.60° |
| 0.10 | -71.6°±0.32° | 71.6°± 0.33° | 143.2°±0.46° |
| 0.20 | -75.4°±0.30° | 75.4°±0.32° | 150.8°±0.44° |
| 0.50 | -80.4°±0.28° | 80.4°±0.26° | 160.8°±0.39° |
| 1.00 | -78.2°±0.26° | 78.2°±0.25° | 156.4°±0.36° |

# Analytical Model for Calibration Curves

The calibration curves were obtained from Analytical Solution Eq. (3). The permittivity of Au was modeled as $\varepsilon_m = 1 - \omega_p^2/\omega^2$ based on Drude model, where $\omega_p$ is the plasma frequency of gold ($\omega_p = 13.8 \times 10^{15}$, Novotny & Hecht, Principle of Nano-optics, pp 369-413, Cambridge University Press. 2012), $\omega$ is the frequency of the incident light. The relationship between the azimuthal angle $\varphi$ and the surrounding refractive index can be obtained using the formula

$$\frac{2\pi}{\lambda} n_d \sin\alpha \pm \frac{2\pi}{\lambda} \sqrt{\frac{\left[1 - \frac{\omega_p^2}{\left(\frac{2\pi c}{\lambda}\right)^2}\right] * n_d^2}{\left[1 - \frac{\omega_p^2}{\left(\frac{2\pi c}{\lambda}\right)^2}\right] + n_d^2}} = \frac{2m\pi}{d\cos\varphi + \sqrt{(-d^2 + 2\Delta r^2 + d^2\cos 2\varphi)/2}} \quad [S5]$$

, where $\lambda$ is the wavelength of the incident light, $\alpha$ is the incident angle, $n_d$ is the refractive index of the surrounding dielectric medium and m is the resonant order, $\varphi$ is the azimuthal angle, d and $\Delta r$ are the parameters for the PDG structure, c is the speed of light in vacuum. The shaded areas along the curves are the analytically predicted range of in-coupling angle due to the finite bandwidth of the wavelength of the illumination determined by the bandwidth of the bandpass filter (~40 nm).

# Small-range index sensing data from a PDG with d= 20 nm and Δr = 390 nm

In this work, we have fabricated two PDG index sensors for the small-range index sensing experiment, i.e. the index sensing of water-ethanol mixture. One PDG is designed to have d = 40 nm and Δr = 390 nm and the other d = 20 nm and Δr = 390 nm. Index sensing results from the former (d = 40 nm and Δr = 390 nm) is presented in the main text (Fig. 4 and Table 1). Here, we also provide the results from the latter (d = 20 nm and Δr = 390 nm) for reference. The original reflection images and intensity profiles are shown in Fig. S2. All fitting parameters are provided in Table S4. The obtained open angles and corresponding refractive indexes are given in Table S5. Both designs are sensitive enough to measure the nonlinear index change of water-ethanol mixture.

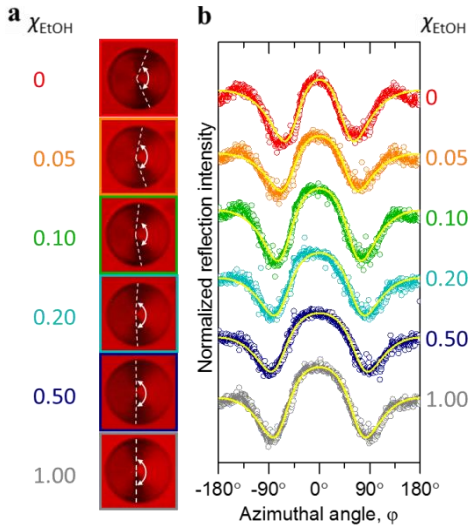

**Figure S2.** (a) Reflection images of the PDG index sensor illuminated by a near-normally incident white light filtered by a narrow band filter centered at 632.8 nm. Shown from the top to the bottom rows are the reflection images of the PDG index sensor covered with water-ethanol mixture with molar ratio of ethanol $\chi_{EtOH}$ = 0, 0.05, 0.10, 0.20, 0.50 and 1.00. The dashed lines mark the azimuthal angles of the SP resonant bands with m = -1 and the position of the 0° reference. (b) Intensity angle distribution extracted from the reflection images in (a) at corresponding molar ratio of ethanol. Solid traces are obtained from fitting the experimental data with an analytical model based on Fano resonance model.

**Table S4.** Parameters used for fitting reflection intensity profiles for PDG with the condition of d = 20 nm and $\Delta r$ = 390 nm.

| \multicolumn{2}{c}{Fitting parameters used for PDG with d= 20 nm and $\Delta r$ = 390 nm} | | | | | | |
|---|---|---|---|---|---|---|---|
| parameter | | $\chi_{EtOH} = 0$ | $\chi_{EtOH} = 0.05$ | $\chi_{EtOH} = 0.10$ | $\chi_{EtOH} = 0.20$ | $\chi_{EtOH} = 0.50$ | $\chi_{EtOH} = 1.00$ |
| $P_m$ | | 403.9559 | 401.6211 | 399.0951 | 397.3678 | 394.9006 | 396.9062 |
| $w_m$ | | 10.74348 | 11.20124 | 12.12631 | 13.16701 | 13.75906 | 13.00903 |
| $A_m$ | $m = -1$ | 1 | 1 | 1 | 1 | 1 | 1 |
| $q_m$ | | 0.36294 | 0.27089 | 0.29892 | 0.25055 | 0.20088 | 0.28071 |
| $b_m$ | | 0.23821 | 0.43128 | 0.16748 | 0.32002 | 0.34875 | 0.26477 |
| $\Delta r$ | | 390 | 390 | 390 | 390 | 390 | 390 |
| $d$ | | 20 | 20 | 20 | 20 | 20 | 20 |
| $\eta$ | | 0 | 0 | 0 | 0 | 0 | 0 |
| $\gamma$ | | 0 | 0 | 0 | 0 | 0 | 0 |
| $y_0$ | | 0.98143 | 0.89311 | 1.00961 | 0.90805 | 0.89882 | 0.9144 |

**Table S5.** Summary of the open angles ($2\varphi$) and the corresponding refractive indexes ($n_d$) of the water-ethanol mixture solutions with $d$ = 20 nm and $\Delta r$ = 390 nm.

| $\chi_{EtOH}$ | $2\varphi$ | $n_d$ |
|---|---|---|
| 0 | 123.4±0.50 | 1.3376±0.00046 |
| 0.05 | 138.2±0.63 | 1.3446±0.00062 |
| 0.10 | 149.2±0.35 | 1.3501±0.00036 |
| 0.20 | 161.0±0.45 | 1.3561±0.00047 |
| 0.50 | 172.4±0.53 | 1.3621±0.00055 |
| 1.00 | 165.0±0.35 | 1.3582±0.00036 |